\documentclass[conference,10pt]{IEEEtran}
\IEEEoverridecommandlockouts
\usepackage{cite}
\usepackage{balance}
\usepackage{amsmath,amssymb,amsfonts}
\usepackage{algorithmic}
\usepackage{graphicx}
\usepackage{textcomp}
\usepackage{multirow,booktabs}
\usepackage[table]{xcolor}
\def\BibTeX{{\rm B\kern-.05em{\sc i\kern-.025em b}\kern-.08em
    T\kern-.1667em\lower.7ex\hbox{E}\kern-.125emX}}
\begin{document}

\title{Efficient Scaling for LLM-based ASR\\
}
\author{
    \IEEEauthorblockN{
	\textit{Bingshen Mu}\IEEEauthorrefmark{4}, 
	\textit{Yiwen Shao}\IEEEauthorrefmark{4},
        \textit{Kun Wei}\IEEEauthorrefmark{4},
        \textit{Dong Yu}\IEEEauthorrefmark{4},
        \textit{Lei Xie}
  }
        \IEEEauthorblockA{\IEEEauthorrefmark{4}Tencent AI Lab
}}

\maketitle

\begin{abstract}
Large language model (LLM)-based automatic speech recognition (ASR) achieves strong performance but often incurs high computational costs. This work investigates how to obtain the best LLM-ASR performance efficiently. Through comprehensive and controlled experiments, we find that pretraining the speech encoder before integrating it with the LLM leads to significantly better scaling efficiency than the standard practice of joint post-training of LLM-ASR. Based on this insight, we propose a new multi-stage LLM-ASR training strategy, EFIN: Encoder First Integration. Among all training strategies evaluated, EFIN consistently delivers better performance (relative to 21.1\% CERR) with significantly lower computation budgets (49.9\% FLOPs). Furthermore, we derive a scaling law that approximates ASR error rates as a computation function, providing practical guidance for LLM-ASR scaling.
\end{abstract}


\section{Introduction}\label{Introduction}
Since the rise of neural networks (NNs) and deep learning in the 2010s, automatic speech recognition (ASR) has evolved from hybrid frameworks~\cite{graves2006connectionist, hinton2012deep, miao2015eesen} that relied solely on NN-based acoustic models to end-to-end (E2E) frameworks~\cite{bahdanau2016end,chan2016listen, dong2020cif, graves2012sequence, watanabe2017hybrid}, in which the entire NN model is trained to output transcription text directly.
Despite significant progress in speech recognition accuracy measured by word error rate (WER) or character error rate (CER), a considerable number of errors persist~\cite{mu2024automatic, mu2024mmger, mu2025hdmole}.
Specifically, current E2E ASR frameworks struggle to efficiently leverage rich commonsense knowledge and perform contextual reasoning during the speech recognition process, making them inevitably reliant on complicated fusion strategies with external language models (LMs). 
With the rapid advancement of large language models (LLMs)~\cite{achiam2023gpt, touvron2023llama, touvron2023llama2, grattafiori2024llama, bai2023qwen, yang2025qwen3}, the potential of artificial intelligence continues to grow. 
Substantial research has focused on exploring the potential of LLMs in various fields, particularly in ASR.
The extensive text knowledge and contextual reasoning capabilities stored in LLMs make them potential components for providing semantic guidance to ASR.
Recent developments in combining LLM with ASR have led to outstanding performance, with the paradigm of connecting a speech encoder with an LLM through a projection layer becoming the prevailing framework for LLM-based ASR (LLM-ASR)~\cite{tang2023salmonn, rubenstein2023audiopalm, huang2024audiogpt, chu2023qwen, chu2024qwen2, ma2024embarrassingly, geng2024unveiling, mu2025mixture}.
However, the large number of parameters in LLM-ASR requires an enormous computational budget for training.
Therefore, the research question of our work can be described as: \textit{How to efficiently train LLM-ASR to achieve optimal performance under a given computational budget?}

Previous studies have explored various training strategies for LLM-ASR~\cite{ma2024embarrassingly,geng2024unveiling,shi2024advancing}. Still, they typically focus on optimizing only a specific component or jointly optimizing multiple modules within the entire LLM-ASR framework, resulting in a substantial computational budget.
Moreover, LLM-ASR typically utilizes the encoder from an existing pretrained ASR model (e.g., Whisper~\cite{radford2023robust}) and keeps it frozen during training.
In this work, we present a crucial insight: \textit{pretraining the speech encoder before integrating it with the LLM leads to significantly better scaling efficiency than the standard practice of joint post-training of LLM-ASR.}
Specifically, since the parameter of the ASR model is much smaller than that of LLM-ASR, independently training a ASR model with high recognition accuracy is more cost-effective. 
Additionally, the encoder of this ASR model possesses better speech feature extraction capabilities, which significantly contributes to improving the recognition performance of LLM-ASR.
Therefore, given a pretrained speech encoder and a pretrained LLM as backbones, along with new in-domain data for post-training, we introduce a three-stage training strategy for LLM-ASR, named \textbf{EFIN}: \textbf{E}ncoder \textbf{F}irst \textbf{IN}tegration.

\begin{itemize}
    \item \textbf{Stage 1:} We fine-tune the speech encoder independently using its original architecture and objective.
    
    \item \textbf{Stage 2:} We freeze both the fine-tuned encoder and the pretrained LLM, and train only the projection layer to preliminary convergence.
    
    \item \textbf{Stage 3:} We unfreeze the projection layer and the LLM, and jointly train them toward final convergence. To further reduce resource requirements, we apply Low-Rank Adaptation (LoRA)~\cite{hu2022lora} to the LLM during this stage.
\end{itemize}

Furthermore, we investigate the scaling properties of the proposed training strategy EFIN to accurately predict the speech recognition performance of LLM-ASR under a fixed computational budget. By utilizing different fine-tuned Whisper encoders, we train the projection layer and LLM on top of them (i.e., the last two stages) with datasets ranging from 2K to 10K hours to derive a neural scaling law. This law reveals that the ASR error rate follows a power-law relationship with the total computational cost (FLOPs).

We also compare the scaling behaviors of various training strategies and observe that our proposed strategy EFIN consistently achieves better recognition performance under the same computational budget. 
Moreover, when the speech encoder is pretrained with a larger model capacity and more extensive data, the downstream performance of LLM-ASR further improves. These findings validate our core insight and suggest that pretraining or fine-tuning a better speech encoder with greater computational resources yields a more favorable final LLM-ASR performance.

In general, our contributions are summarized as follows:
\begin{itemize}
    \item We conduct comprehensive experiments to evaluate multiple stages of existing LLM-ASR training strategies and identify the most computationally efficient approach. We demonstrate that pretraining or fine-tuning a speech encoder independently (i.e., from its original architecture) yields higher efficiency and better final performance than conventional joint encoder and LLM training. Building on this insight, we propose a three-stage LLM-ASR training strategy named \textbf{EFIN}.
    
    \item We derive a power-law relationship between ASR error rate and computational budget (FLOPs) for \textbf{EFIN}, based on training with datasets ranging from 2K to 10K hours. This neural scaling law empirically predicts LLM-ASR performance when scaling up.
\end{itemize}

\begin{figure}[t]
  \centering
  \includegraphics[width=1\linewidth]{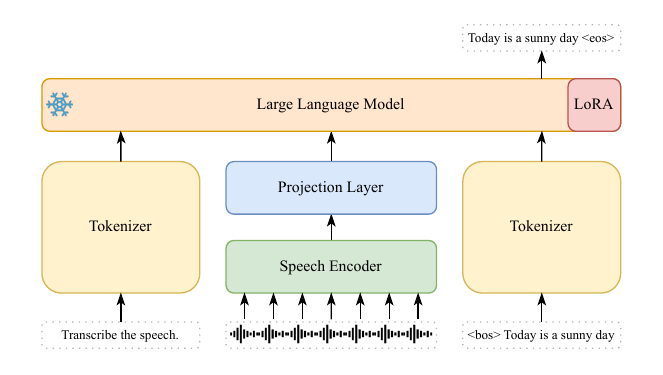}
  \caption{Overview of the LLM-ASR. There are three main components: the speech encoder, the projection layer, and the LLM. LoRA is used for parameter-efficient fine-tuning of the LLM.}
  \label{fig:fig1}
\end{figure}
\section{Related Work}

\subsection{LLM-ASR and Training Strategies}
As shown in Figure~\ref{fig:fig1}, LLM-ASR primarily consists of three components: the speech encoder, the projection layer, and the LLM.
For each speech sample, the prompt used for transcribing (i.e., transcribe the speech), the speech feature sequence, and the corresponding transcription are denoted as $P$, $S$, and $T$, respectively.
The prompt and transcription are tokenized using the tokenizer and then passed through the embedding layer of the LLM to obtain the embedding sequences $E_p$ and $E_t$, which can be denoted as:
\begin{gather}
    E_p = \text{Embedding}(\text{Tokenizer}(P)),\\
    E_t = \text{Embedding}(\text{Tokenizer}(T)).
\end{gather}
Then, the speech feature sequence $S$ is processed by the speech encoder and the projection layer to obtain a feature sequence $E_s$ aligned with the LLM text modality, which can be expressed as follows:
\begin{equation}
    E_s = \text{Projection}(\text{Encoder}(S)).
\end{equation}
Finally, the LLM auto-regressively predicts the transcription $Y$ based on the concatenated feature sequences $E_p$, $E_s$, and $E_t$, which can be formulated as:
\begin{equation}
    Y = \text{LLM}(\text{Concat}(E_p,E_s,E_t)).
\end{equation}

Based on the above LLM-ASR architecture, numerous studies have explored various training strategies.
LauraGPT~\cite{du2023lauragpt} connects a modified speech encoder to the LLM for end-to-end training across various speech and audio tasks, performing full-parameter fine-tuning.
SLAM-ASR~\cite{ma2024embarrassingly} trains only the linear projection layer and achieves remarkable performance on the LibriSpeech~\cite{panayotov2015librispeech} dataset by aligning the speech encoder with the LLM.
Geng et al.~\cite{geng2024unveiling} achieve outstanding performance on multiple open-source Chinese datasets using a three-stage training strategy, which involves separately training the projection layer, the speech encoder, and the LLM with LoRA.
SALMONN~\cite{tang2023salmonn} trains the projection layer and the LLM with LoRA, enabling it to perform multiple speech and audio tasks.
Qwen-Audio~\cite{chu2023qwen} fine-tunes the speech encoder and projection layer, using the loss from the frozen LLM output for backpropagation optimization, transforming it into a universal audio model.
FireRedASR-LLM~\cite{xu2025fireredasr} initializes the speech encoder in the LLM-ASR with a pretrained speech encoder and jointly updates the parameters of the speech encoder, the projection layer, and the LLM with LoRA during training.
\subsection{Neural Scaling Laws}
Previous studies have shown that the performance of Transformer-based~\cite{vaswani2017attention} models at scale can be empirically predicted using three fundamental variables: the model size $N$, the training data size $D$, and the computational budget $B$~\cite{hestness2017deep, ghorbani2021scaling, fernandes2023scaling}.
This can be generalized by modeling the variation in cross-entropy loss $L$ as each variable is independently varied:
\begin{equation}
    L(x) = L_\infty + \beta _xx^{\alpha _x},
\end{equation}
where $x\in (N,D,B)$, $L(x)$ represents the reducible loss that follows a power-scaling law, while $L_{\infty}$ denotes the irreducible loss.
$\beta_x$ and $\alpha_x$ are empirically determined power-law variables.
Varying the value of $x$ allows estimation of the scaling behavior under different settings.

Some researchers have also explored the application of scaling laws in speech tasks.
Gu et al.~\cite{gu2023scaling} evaluate the scaling laws of language model re-scoring and find that CER can also be modeled as a power-law function of $x$, which can be expressed as:
\begin{equation}
    \text{CER}(x) = \beta _xx^{\alpha _x}. \label{eq_scale}
\end{equation}
Droppo \& Elibol~\cite{droppo2021scaling} demonstrate that acoustic models trained with an auto-predictive coding loss behave as if they are subject to similar scaling laws.
Cuervo \& Marxer~\cite{cuervo2024scaling} devise the scaling laws for speech language models.
OWSL~\cite{chen2025owls} investigates the scaling laws of large-scale multilingual speech recognition and translation models, demonstrating that the effects of scaling parameters, training data, and computing can lead to reasonable direct predictions of downstream speech recognition and translation performance.
\section{Efficient Training Strategy}\label{sec3}
In this section, we introduce an LLM-ASR training strategy named EFIN according to the insight presented in Section~\ref{Introduction} and highlight its advantages through comparisons with other training strategies.
\subsection{Experimental Setup}\label{sec3a}
\textbf{LLM-ASR Structure.} We follow the LLM-ASR architecture illustrated in Figure~\ref{fig:fig1}. The speech encoder is the Whisper-medium\footnote{https://huggingface.co/openai/whisper-medium} encoder, the projection layer consists of two linear layers with a ReLU activation function, and the LLM is Qwen2.5-7B-Instruct\footnote{https://huggingface.co/Qwen/Qwen2.5-7B-Instruct}. LoRA is applied with a rank of 64 and an alpha of 16 and is integrated into seven modules of each LLM layer including q\_proj, k\_proj, v\_proj, o\_proj, up\_proj, gate\_proj, and down\_proj.

\textbf{Datasets.}
We conduct comparative experiments between different training strategies using the 10K-hour open-source Chinese dataset WenetSpeech~\cite{zhang2022wenetspeech} with high-quality annotations.
For each stage in the training strategies, we use the entire WenetSpeech dataset.

\subsection{Multi-stage Training}

In LLM-ASR, training typically proceeds in multiple stages, differentiated by which modules are involved. We summarize the main stages as follows:

\begin{itemize}
    \item \textbf{Alignment Stage:} Train only the projection layer to align the speech encoder representations with the LLM's text embedding space.
    
    \item \textbf{LLM Adaptation Stage:} Train the projection layer together with the LLM to further adapt the LLM to speech-domain data. LoRA can optionally be applied to the LLM for efficient fine-tuning.
    
    \item \textbf{Full Joint Training Stage:} All three modules—speech encoder, projection layer, and LLM—are jointly optimized until convergence.
\end{itemize}

\subsection{Baseline Training Strategies}\label{baseline}
We progressively incorporated these three stages into training to establish multiple strategies for comparison.
\begin{itemize}
    \item \textbf{Strategy-1:} (1) Alignment Stage only. This is equivalent to SLAM-ASR~\cite{ma2024embarrassingly}.
    \item \textbf{Strategy-2:} (1) Alignment Stage; (2) LLM Adaptation Stage with {LoRA}. This strategy is consistent with the icefall recipe\footnote{https://github.com/k2-fsa/icefall/tree/master/egs/speech\_llm/ASR\_LLM}.
    \item \textbf{Strategy-3:} (1) Alignment Stage; (2) LLM Adaptation Stage with LoRA; (3) Full Joint Training Stage.
\end{itemize}

\begin{figure}[t]
  \centering
  \includegraphics[width=1\linewidth]{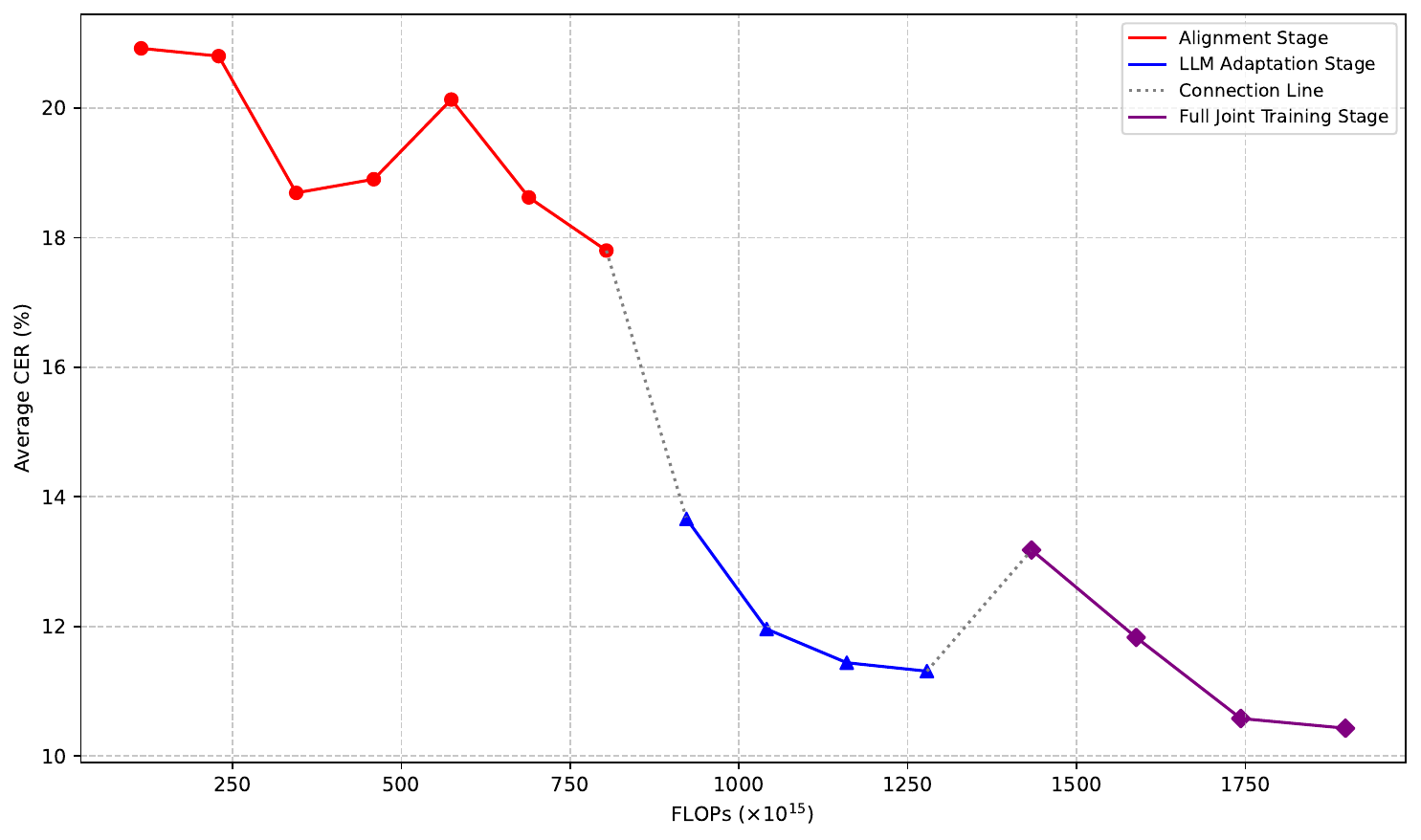}
  \caption{The average CER and FLOPs at each checkpoint for Strategy-1 (SLAM-ASR), Strategy-2 and Strategy-3. \textbf{Strategy-1 (SLAM-ASR):} the red segment only; \textbf{Strategy-2:} the red and blue segments; \textbf{Strategy-3:} the red, blue, and purple segments.}
  \label{fig:fig2}
\end{figure}

As shown in Figure~\ref{fig:fig2} and Table~\ref{table1}, we find that Strategy-1 (SLAM-ASR), which trains only the projection layer, contributes limited improvement to the speech recognition performance of LLM-ASR.
Loading the parameters of the projection layer and then training it together with the LLM using LoRA yields significant performance gains (Strategy-2).
Building upon this, jointly training the speech encoder, projection layer, and LLM using LoRA yields only marginal performance gains while incurring significantly higher FLOPs consumption (Strategy-3).

\begin{table}[h]
\caption{CER (\%) and total FLOPs ($\times 10^{15}$) for all training strategies. ``CER'' refers to the character error rate of LLM-ASR on two test sets.}
\label{table1}
\centering
\scalebox{0.90}{
\begin{tabular}{clccc}
\toprule
\multirow{2}{*}{} & \multirow{2}{*}{Strategy} & \multicolumn{2}{c}{CER (\%)} & \multirow{2}{*}{FLOPs ($\times 10^{15}$)} \\
\cmidrule(lr){3-4}
& & TEST-MEETING & TEST-NET & \\
\midrule
\multirow{3}{*}{\textit{Baseline}} 
& Strategy-1 & 18.76 & 16.84 & \textbf{803.77} \\
& Strategy-2 & 12.20 & 10.42 & 1278.39 \\
& Strategy-3 & 12.37 & 8.48  & 1898.16 \\
\midrule
\multirow{3}{*}{\textit{EFIN}} 
& Strategy-4 & 11.33 & 8.38  & 1162.58 \\
& Strategy-5 & \textbf{9.54}  & \textbf{7.01}  & 1637.20 \\
& Strategy-6 & 12.23 & 8.58  & 2102.03 \\
\bottomrule
\end{tabular}}
\end{table}

\begin{figure}[h]
  \centering
  \includegraphics[width=1\linewidth]{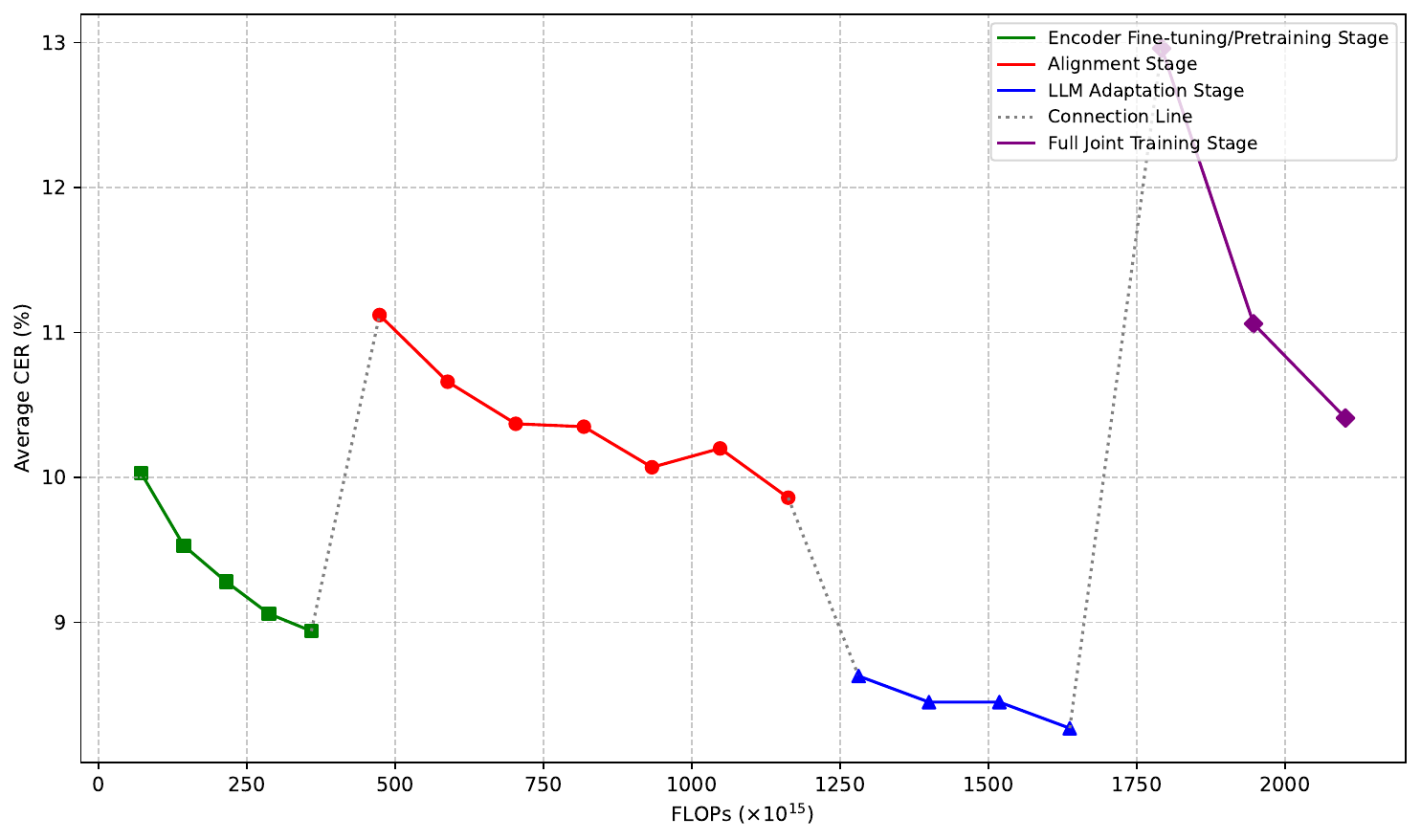}
  \caption{The average CER and FLOPs at each checkpoint for Strategy-4, Strategy-5 and Strategy-6. \textbf{Strategy-4:} the green and red segments; \textbf{Strategy-5:} the green, red and blue segments; \textbf{Strategy-6:} the green, red, blue, and purple segments.}
  \label{fig:fig3}
\end{figure}

\subsection{EFIN: Encoder First Integration}

Although computationally inefficient, the marginal improvement from the Full Joint Training Stage suggests that optimizing the speech encoder still holds potential for performance gains. This motivates us to investigate whether it is beneficial to introduce an additional stage \textit{before} the Alignment Stage to fine-tune the speech encoder independently using its original architecture and objective. We refer to this strategy as \textbf{EFIN:} \textbf{E}ncoder \textbf{F}irst \textbf{IN}tegration.

Unlike the Full Joint Training Stage—which incurs high computational costs with diminishing returns in optimizing the encoder—our approach fine-tunes the encoder separately in a much more efficient and straightforward manner. The key question is \textit{whether the benefits from this encoder fine-tuning will persist through subsequent training stages.}

To answer this, we progressively incorporate stages into the training pipeline, as described in Section~\ref{baseline}:

\begin{itemize}
    \item \textbf{Strategy-4:} (1) Encoder Fine-tuning/Pretraining Stage; (2) Alignment Stage.
    \item \textbf{Strategy-5:} (1) Encoder Fine-tuning/Pretraining Stage; (2) Alignment Stage; (3) LLM Adaptation Stage.
    \item \textbf{Strategy-6:} (1) Encoder Fine-tuning/Pretraining Stage; (2) Alignment Stage; (3) LLM Adaptation Stage; (4) Full Joint Training Stage.
\end{itemize}

Both Table~\ref{table1} and Figure~\ref{fig:fig3} show that \textbf{EFIN} significantly improves ASR performance over baseline strategies while incurring only a modest increase in computational cost (e.g., Strategy-5 vs. Strategy-2). A notable finding is that once the training begins with the Encoder Fine-tuning Stage, adding a Full Joint Training Stage is no longer necessary and may even degrade performance. This is evidenced by the drop in accuracy from Strategy-5 to Strategy-6.

From Figure~\ref{fig:fig3}, one might speculate that with additional computational resources (e.g., more training steps or larger datasets), Strategy-6 could eventually catch up to Strategy-5. However, we argue that this is not worthwhile given its significantly higher cost due to Full Joint Training Stage. Therefore, we finalize \textbf{EFIN} as a three-stage training strategy (i.e., Strategy-5), omitting the Full Joint Training Stage for better efficiency-performance trade-off.

Furthermore, comparing Strategy-3 and Strategy-5 reveals an important insight: \textit{pretraining the encoder is more compute-efficient and more effective}. With only 86\% of the computational budget of Strategy-3, Strategy-5 achieves a relative CER reduction of 22.8\% on TEST-MEETING and 17.3\% on TEST-NET, respectively.

\subsection{Towards More Efficient EFIN}\label{sec:sec_efin}

\begin{figure}[h]
  \centering
  \includegraphics[width=1\linewidth]{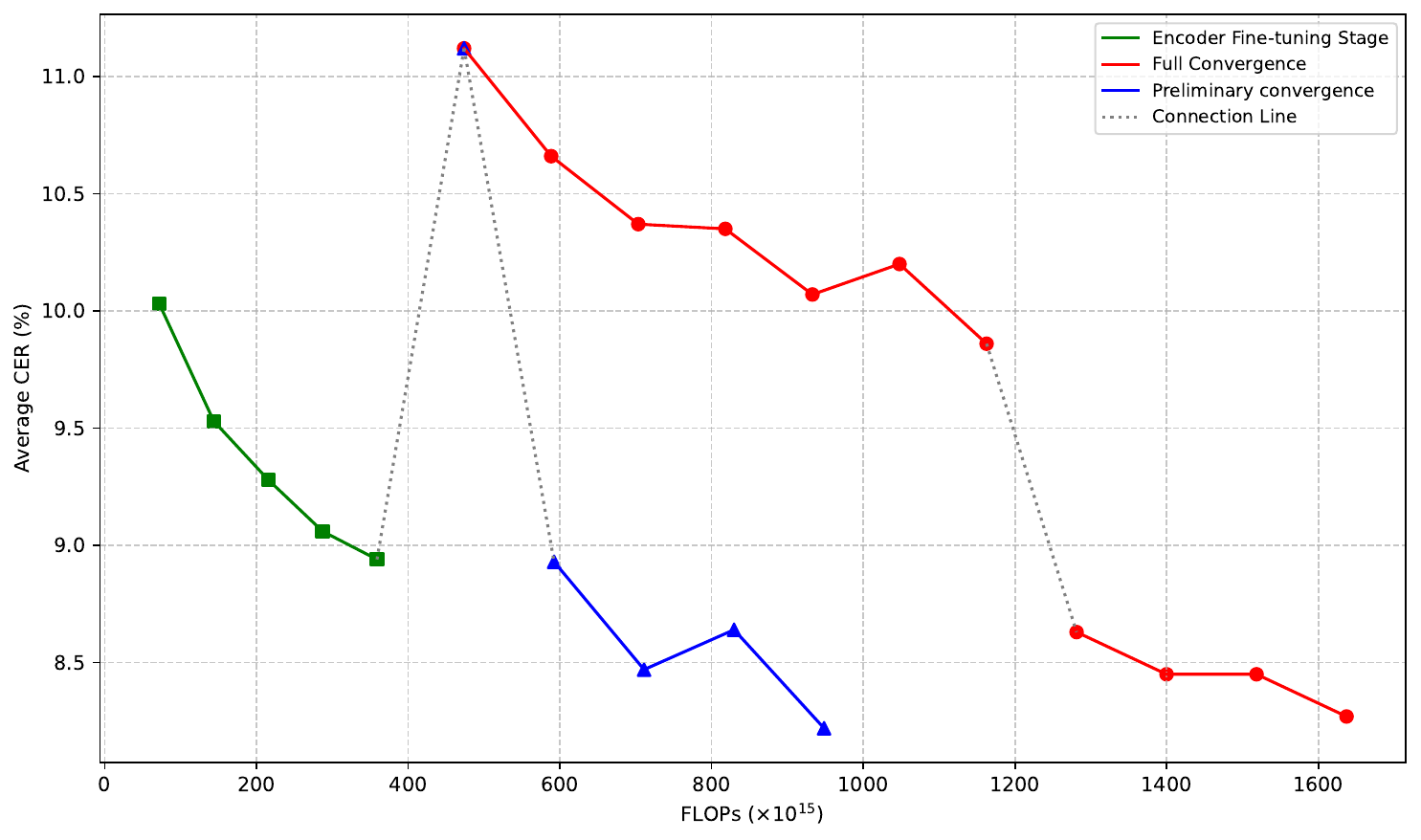}
  \caption{The average CER and FLOPs at each checkpoint for different levels of convergence in the second stage of Strategy-5. \textbf{Green:} Encoder Fine-tuning Stage; \textbf{Blue:} Alignment Stage with preliminary convergence and LLM Adaptation Stage; \textbf{Red:} Alignment Stage with full convergence and LLM Adaptation Stage.}
  \label{fig:fig4}
\end{figure}

While EFIN already achieves strong performance under a constrained computational budget, we further explore whether its efficiency can be improved. In particular, we revisit the Alignment Stage, which is designed to bring the projection layer into alignment with the LLM. As shown in Figure~\ref{fig:fig3}, this stage (highlighted in red) consumes a relatively large number of FLOPs and exhibits slow convergence.

Given that the subsequent LLM Adaptation Stage also updates the projection layer, we hypothesize that \textit{full convergence during the Alignment Stage may be unnecessary}. Instead, guiding the projection layer toward \textit{preliminary alignment} may be sufficient to support effective downstream optimization. This opens up the possibility of reducing compute usage by shortening the Alignment Stage without degrading overall speech recognition performance.
\begin{table}[h]
\caption{CER (\%) and total FLOPs ($\times10^{15}$) for different levels of convergence in the second stage of Strategy-5. ``CER'' refers to the character error rate of LLM-ASR on two test sets.}
\label{table2}
\centering
\scalebox{0.90}{
\begin{tabular}{lccc}
\toprule
                        & \multicolumn{2}{c}{CER (\%)} & \multirow{2}{*}{FLOPs ($\times10^{15}$)} \\ \cmidrule{2-3}
                        & TEST-MEETING        & TEST-NET         &                                                 \\ \midrule
Full convergence       & 9.54           & 7.01        & 1637.20                                         \\
Preliminary convergence & \textbf{9.45}           & \textbf{7.00}        & \textbf{948.26}                                          \\ \bottomrule
\end{tabular}}
\end{table}

To validate this hypothesis, we conduct experiments to examine whether Strategy-5 requires the Alignment Stage to reach full convergence. Table~\ref{table2} presents the final CER and FLOPs for different levels of convergence in the second stage of Strategy-5. We find that training the projection layer only to a preliminary level of convergence before proceeding to the next stage reduces FLOPs by 42.1\% and leads to a slight improvement in ASR performance. This suggests that full convergence during the Alignment Stage is unnecessary and may even be suboptimal in practice. Furthermore, Figure~\ref{fig:fig4} illustrates the CER and FLOPs at each checkpoint for different levels of convergence for the Alignment Stage of Strategy-5. It shows that the subsequent training process converges significantly faster with only preliminary convergence.

As a result, we finalize our proposed \textbf{EFIN} into its most efficient and effective form:
\begin{itemize}
    \item \textbf{Stage 1:} Encoder Fine-tuning/Pretraining Stage.
    \item \textbf{Stage 2:} Alignment Stage with preliminary convergence.
    \item \textbf{Stage 3:} LLM Adaptation Stage.
\end{itemize}

\section{Scaling Behavior of Various Strategies}\label{sec4}

In this section, we conduct comprehensive experiments on all three baseline and three EFIN training strategies to compare their scaling behaviors under different computational budgets.

\subsection{Experimental Setup}\label{sec:exp}

The overall LLM-ASR structure remains consistent with that described in Section~\ref{sec3a}. For all EFIN strategies that involve fine-tuning or pretraining the speech encoder, we fix the encoder to one that has been fine-tuned on the full 10K-hour WenetSpeech dataset. This allows us to isolate and analyze the scaling behavior of the stages involving the LLM.

For the remaining training stages (i.e., Alignment, LLM Adaptation, and Full Joint Training), we vary the data scale using subsets of 2K, 5K, 8K, and 10K hours randomly sampled from WenetSpeech.

\begin{table}[h]
\caption{CER (\%) and total FLOPs ($\times10^{15}$) for the six training strategies under different data scale. ``AVG'' is the mean of TEST-MEETING and TEST-NET.}
\label{table3}
\centering
\begin{tabular}{lcccc}
\toprule
 & \multicolumn{3}{c}{CER (\%)} & \multirow{2}{*}{FLOPs ($\times10^{15}$)} \\ \cmidrule(lr){2-4}
             & MEETING & NET & AVG & \\ \midrule

\rowcolor{gray!20}\multicolumn{5}{c}{\textbf{\textit{Baseline: Strategy-1 (SLAM-ASR)}}} \\
2,000 Hours  & 22.39 & 19.33 & 20.86 & 160.75 \\
5,000 Hours  & 22.66 & 18.54 & 20.60 & 401.88 \\
8,000 Hours  & 19.47 & 17.34 & 18.41 & 643.01 \\
10,000 Hours & 18.76 & 16.84 & 17.80 & 803.77 \\ \midrule

\rowcolor{gray!20}\multicolumn{5}{c}{\textbf{\textit{Baseline: Strategy-2}}} \\
2,000 Hours  & 14.95 & 12.75 & 13.85 & 255.68 \\
5,000 Hours  & 13.06 & 11.18 & 12.12 & 639.19 \\
8,000 Hours  & 12.19 & 10.50 & 11.35 & 1022.71 \\
10,000 Hours & 12.20 & 10.42 & 11.31 & 1278.39 \\ \midrule

\rowcolor{gray!20}\multicolumn{5}{c}{\textbf{\textit{Baseline: Strategy-3}}} \\
2,000 Hours  & 19.22 & 12.34 & 15.78 & 379.63 \\
5,000 Hours  & 14.47 & 9.69  & 12.08 & 949.08 \\
8,000 Hours  & 13.80 & 9.48  & 11.64 & 1518.53 \\
10,000 Hours & 12.37 & 8.48  & 10.43 & 1898.16 \\ \midrule

\rowcolor{gray!20}\multicolumn{5}{c}{\textbf{\textit{EFIN: Strategy-4}}} \\
2,000 Hours  & 12.57 & 9.49  & 11.03 & 519.57 \\
5,000 Hours  & 12.04 & 8.87  & 10.46 & 760.69 \\
8,000 Hours  & 11.30 & 8.73  & 10.02 & 1001.83 \\
10,000 Hours & 11.33 & 8.38  & 9.86  & 1162.58 \\ \midrule

\rowcolor{gray!20}\multicolumn{5}{c}{\textbf{\textit{EFIN: Strategy-5 preliminary convergence (proposed best)}}} \\
2,000 Hours  & 10.77 & 7.86  & 9.32  & 476.70 \\
5,000 Hours  & 9.81  & 7.45  & 8.63  & 653.54 \\
8,000 Hours  & 9.63  & 7.07  & 8.35  & 830.37 \\
10,000 Hours & \textbf{9.45} & \textbf{7.00} & \textbf{8.23} & 948.26 \\ \midrule

\rowcolor{gray!20}\multicolumn{5}{c}{\textbf{\textit{EFIN: Strategy-6}}} \\
2,000 Hours  & 18.88 & 11.48 & 15.18 & 738.44 \\
5,000 Hours  & 14.53 & 10.02 & 12.28 & 1307.89 \\
8,000 Hours  & 12.43 & 8.57  & 10.50 & 1877.34 \\
10,000 Hours & 12.23 & 8.58  & 10.41 & 2102.03 \\

\bottomrule
\end{tabular}
\end{table}

\begin{figure}[h]
  \centering
  \includegraphics[width=1\linewidth]{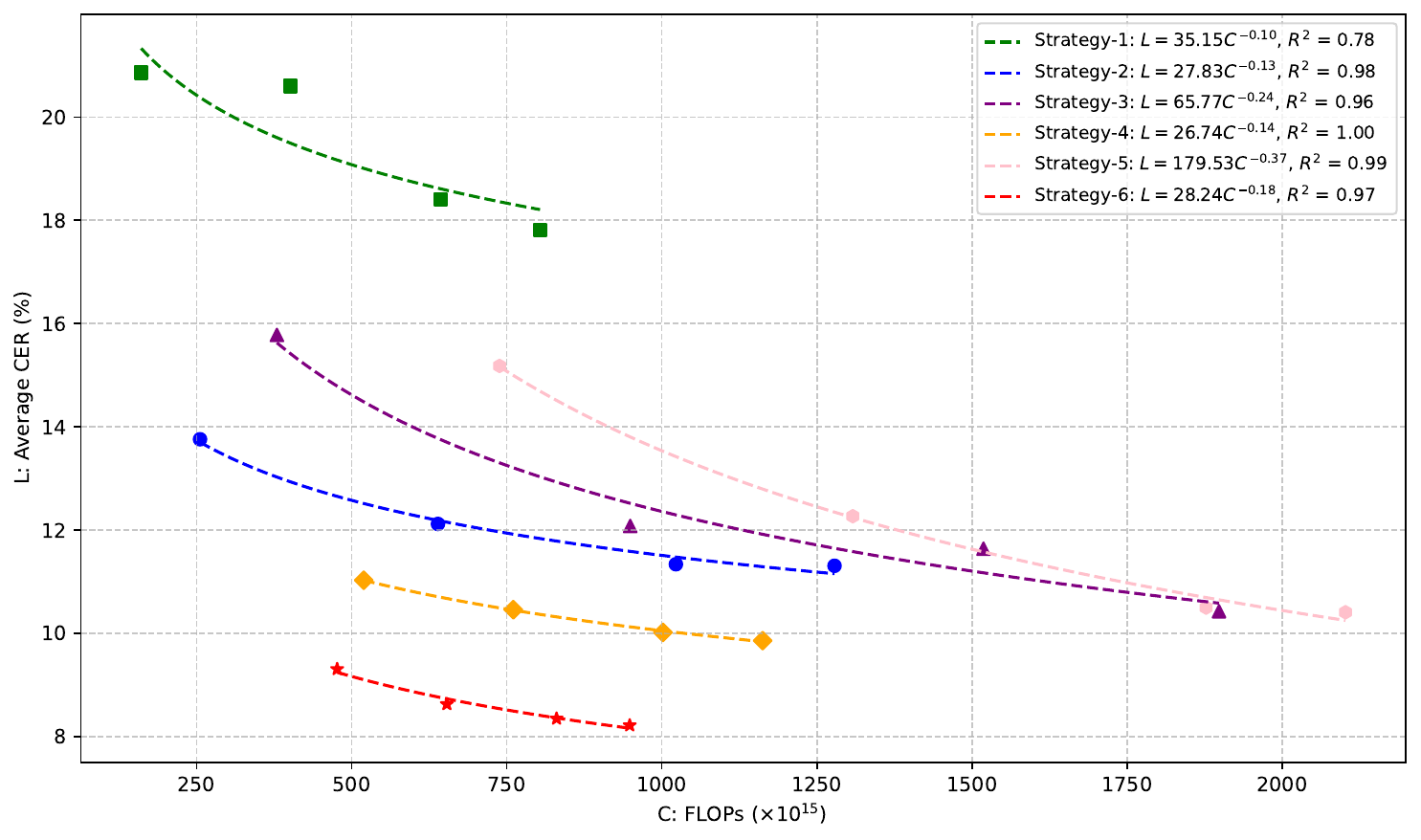}
  \caption{Scaling law curves of multiple training strategies. Each data point represents the final converged CER and corresponding FLOPs for each strategy at a given data scale.}
  \label{fig:fig5}
\end{figure}
\subsection{Experimental Results}
Table~\ref{table3} presents the CER and FLOPs results of the six training strategies under different data scaling settings. Note that the additional FLOPs for fine-tuning the speech encoder in all EFIN strategies are included in the total FLOPs reported.

Compared to the baseline training strategies, our proposed EFIN variants consistently exhibit better scaling behavior—achieving lower CERs with only modest increases in computational cost. Among all evaluated methods, the best-performing strategy is EFIN Strategy-5 (with preliminary convergence), introduced in Section~\ref{sec:sec_efin}, which demonstrates leading ASR performance across all data scales.

In particular, Strategy-5 requires only 18\% more FLOPs than the simplest single-stage Strategy-1 (948.26 vs.\ 803.77) yet achieves a remarkable 53.8\% relative CER reduction (from 17.80\% to 8.23\%). Compared to the strongest baseline, Strategy-3, our best EFIN strategy still achieves a 21.1\% relative CER reduction (from 10.43\% to 8.23\%) while consuming just 49.9\% of the total FLOPs. These results demonstrate that the proposed EFIN training strategy offers both high efficiency and strong effectiveness for LLM-based ASR.

Furthermore, based on the results in Table~\ref{table3}, we plot the scaling law curves of the six training strategies with respect to the training data scale, as shown in Figure~\ref{fig:fig5}. Since the model size is kept constant across all training stages, changes in data scale directly translate to changes in total FLOPs, making the resulting curves approach true compute scaling behavior.

Following the formulation in~\cite{gu2023scaling} and Equation~\ref{eq_scale}, the scaling curves show that the average CER follows a power-law relationship with the total computational budget (FLOPs) for LLM-ASR training. We use the coefficient of determination ($R^2$) to evaluate the goodness of fit between the observed results and the fitted scaling curve.

For our proposed best strategy, EFIN Strategy-5, the scaling behavior follows the power-law:

\begin{equation}
    L = 28.24\,C^{-0.18}, \label{eq7}
\end{equation}
where $L$ denotes the average CER over the TEST-MEETING and TEST-NET, and $C$ represents the computational budget in FLOPs ($\times 10^{15}$). 
This relationship allows us to accurately predict the ASR performance that can be achieved by an LLM-ASR under a given compute budget.

\section{Better Pretrained Speech Encoder}

As demonstrated in Sections~\ref{sec3} and~\ref{sec4}, pretraining or fine-tuning the speech encoder independently is both more compute-efficient and more effective than training it jointly with the LLM in later stages. To further investigate the impact of a well-optimized speech encoder on LLM-ASR performance, we conduct experiments using two pretrained Whisper speech encoders with differing ASR capabilities.

\subsection{Whisper Speech Encoders}

\begin{itemize}
    \item \textbf{Fine-tuned Whisper-medium Encoder:} This encoder is fine-tuned on the full WenetSpeech dataset and is identical to the speech encoder used in our proposed EFIN training strategy described in Sections~\ref{sec3} and~\ref{sec4}.

    \item \textbf{Fine-tuned Whisper-large-v2 Encoder:} This encoder is obtained from an open-source model pretrained on a combination of Chinese ASR datasets, including AISHELL-1~\cite{bu2017aishell}, AISHELL-2~\cite{du2018aishell}, AISHELL-4~\cite{fu2021aishell}, Alimeeting~\cite{yu2022m2met}, KeSpeech~\cite{tang2021kespeech}, and WenetSpeech, totaling 13,906 hours\footnote{https://huggingface.co/yuekai/icefall\_asr\_multi-hans-zh\_whisper}.
\end{itemize}

\subsection{Experimental Setup}
The LLM-ASR structure remains consistent with that described in Section~\ref{sec3a}.
We apply the last two stages of our proposed best EFIN training strategy on the two aforementioned pretrained speech encoders. Same as Section ~\ref{sec:exp}, we randomly select 2K, 5K, and 8K hours of data from the WenetSpeech dataset, as well as the whole 10K hours, to investigate scaling behaviors of LLM-ASR with different pretrained encoders. 
\begin{table}[h]
\caption{CER (\%) and FLOPs ($\times 10^{15}$) for the two whisper speech encoders in LLM-ASR under different data scale.}
\label{table4}
\centering
\begin{tabular}{lccc}
\toprule
             & \multicolumn{2}{c}{CER (\%)} & \multirow{2}{*}{FLOPs ($\times10^{15}$)} \\ \cmidrule{2-3}
             & TEST-MEETING    & TEST-NET   &                                                 \\ \midrule
\rowcolor{gray!20}\multicolumn{4}{c}{\textbf{\textit{Fine-tuned Whisper-medium Encoder}}}                                                                 \\
2,000 Hours  & 10.77           & 7.86       & 476.70                                          \\
5,000 Hours  & 9.81            & 7.45       & 653.54                                          \\
8,000 Hours  & 9.63            & 7.07       & 830.37                                          \\
10,000 Hours & 9.45            & 7.00       & 948.26                                          \\ \midrule
\rowcolor{gray!20}\multicolumn{4}{c}{\textbf{\textit{Fine-tuned Whisper-large-v2 Encoder}}}                                                                 \\
2,000 Hours  & 8.94            & 7.48       & 789.50                                          \\
5,000 Hours  & 8.54            & 7.09       & 1040.57                                         \\
8,000 Hours  & 8.19            & \textbf{6.78}       & 1291.64                                         \\
10,000 Hours & \textbf{7.90}            & 6.81       & 1459.02                                         \\ \bottomrule
\end{tabular}
\end{table}

\begin{figure}[h]
  \centering
  \includegraphics[width=1\linewidth]{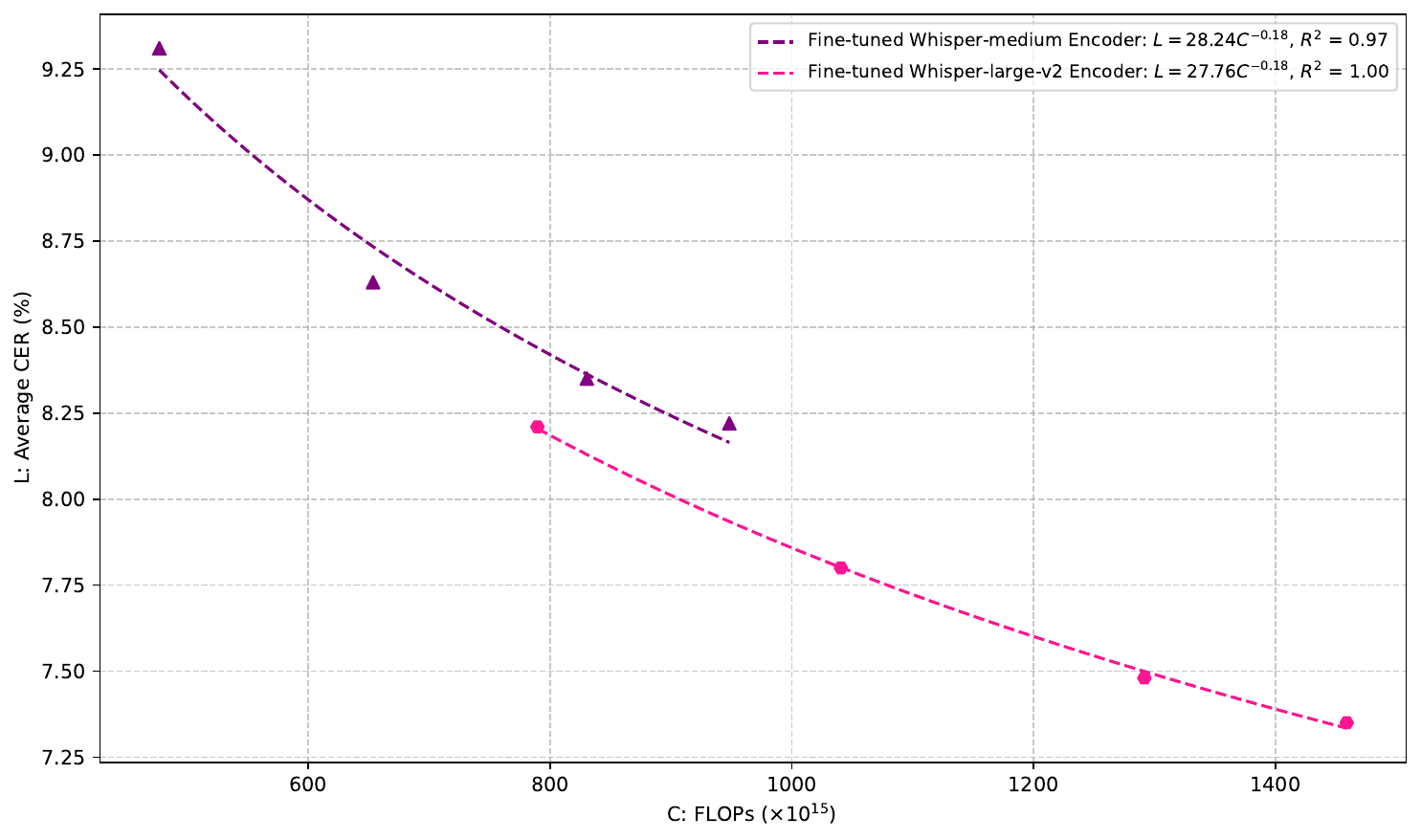}
  \caption{Scaling law curves of the best proposed EFIN training strategy with two pretrained speech encoders. Each data point represents the final converged CER and corresponding FLOPs for each strategy at a given data scale.}
  \label{fig:fig6}
\end{figure}
\subsection{Experimental Results}
As shown in Table~\ref{table4}, with a stronger fine-tuned Whisper-large-v2 speech encoder, we achieve further CER reduction from the previous best number of 9.45\% and 7.00\% to 7.90\% and 6.78\%, respectively. 

Similarly, based on the results in Table~\ref{table4}, we plot the scaling law curves for the two pretrained speech encoders, as shown in Figure~\ref{fig:fig5}. The new power-law scaling relationship for the fine-tuned Whisper-large-v2 encoder is given by:

\begin{equation}
    L = 27.76\,C^{-0.18}. \label{eq8}
\end{equation}

Comparing Equation~(\ref{eq7}) and Equation~(\ref{eq8}), we observe a lower scaling coefficient for the fine-tuned Whisper-large-v2 encoder, indicating improved final LLM-ASR performance under the same computational budget. This result reinforces the importance of strong encoder pretraining: \textit{By investing more computational effort into building a better speech encoder independently, we can achieve superior LLM-ASR performance more efficiently}.

\section{Conclusion}

Integrating ASR with LLMs has become a mainstream trend. However, the substantial computational budget required for training limits the widespread adoption of the LLM-ASR. This work investigates how to train LLM-ASR more efficiently to achieve optimal performance under a constrained computational budget.
Through comprehensive and controlled experiments, we find that pretraining or fine-tuning the speech encoder before integrating it with the LLM yields significantly better scaling efficiency than the standard joint training strategies.
Accordingly, we propose an efficient three-stage training strategy for LLM-ASR, named \textbf{EFIN}, consisting of: (1) pretraining or fine-tuning the speech encoder; (2) training only the projection layer to preliminary convergence; and (3) jointly training the projection layer and the LLM using LoRA.
Furthermore, we derive and analyze the scaling laws for EFIN and other existing strategies. Our results demonstrate that the proposed EFIN strategy consistently outperforms baselines in both computational efficiency and ASR performance. The derived power law also enables accurate prediction of average CER as a function of computational budget when scaling up training.
In future work, we encourage further investigation into the role of well-optimized encoders in other multi-modal LLM settings beyond ASR.

\clearpage
\bibliographystyle{IEEEtran}
\bibliography{refs}

\end{document}